\documentclass[runningheads]{llncs}
\usepackage[T1]{fontenc}
\usepackage{graphicx, xcolor}
\usepackage{booktabs}
\usepackage{amsmath, amssymb, amsfonts}
\usepackage{longtable}
\usepackage{xspace}
\usepackage{array}
\usepackage{hyperref}
\hypersetup{hidelinks}

\newcommand{\TVD}{\mathrm{TVD}}

\newcommand{\normaldist}[2]{\mathcal{N}(#1, #2)}

\newcommand{\citeX}[1]{%
    \textcolor{red}{[%
        \if\relax\detokenize{#1}\relax X\else #1\fi
    ]}%
}

\newcommand{\placeholderbox}[2][]{%
    \fbox{%
        \begin{minipage}[c][5cm][c]{0.8\linewidth}
            \centering
            \vspace{1cm}
            \textbf{[#2]} 
            \if\relax\detokenize{#1}\relax\else \\ #1 \fi 
            \vspace{1cm}
        \end{minipage}%
    }
}

\newcommand{\ppp}{Personalization Policy Probe\xspace}
\newcommand{\pppshort}{P$^3$\xspace}

\begin{document}

\title{Learning Context Matters: Measuring \\ and Diagnosing Personalization Gaps in \\ LLM-Based Instructional Design}

\titlerunning{Learning Context Matters}

\author{Johaun Hatchett\inst{1,2} \and Debshila Basu Mallick\inst{2,3} \and Brittany C. Bradford\inst{2,3} \and Richard G.~Baraniuk \inst{1,2,3}}
\authorrunning{J. Hatchett et al.}
\institute{Rice University, Houston, TX, USA \and SafeInsights \and OpenStax, Houston, TX, USA \\ \email{\{jh152, debshila, bb25, richb\}@rice.edu}}


\maketitle

\begin{abstract}
    The adoption of generative AI in education has accelerated dramatically in recent years, with Large Language Models (LLMs) increasingly integrated into learning environments in the hope of providing personalized support that enhances learner engagement and knowledge retention.
    However, truly personalized support requires access to meaningful Learning Context (LC) regarding who the learner is, what they are trying to understand, and how they are engaging with the material.
    In this paper, we present a framework for measuring and diagnosing how the LC influences instructional strategy selection in LLM-based tutoring systems.
    Using psychometrically grounded synthetic learning contexts and a pedagogically grounded decision space, we compare LLM instructional decisions in context-blind and context-aware conditions and quantify their alignment with the pedagogical judgments of subject matter experts. 
    Our results show that, while providing the LC induces systematic, measurable changes in instructional decisions that move LLM policies closer to the subject matter expert policy, substantial misalignment remains.
    To diagnose this misalignment, we introduce a relevance-impact analysis that reveals which learner characteristics are attended to, ignored, or spuriously influential in LLM instructional decision-making.
    This analysis, conducted in collaboration with subject matter experts, demonstrates that LC materially shapes LLM instructional planning but does not reliably induce pedagogically appropriate personalization. 
    Our results enable principled evaluation of context-aware LLM systems and provide a foundation for improving personalization through learner characteristic prioritization, pedagogical model tuning, and LC engineering.

\end{abstract}

\section{Introduction}

Personalized instruction has long been recognized as more effective than one-size-fits-all approaches in education \cite{bloom1984-twosigma,vanlehn2011-relative}. In human tutoring, instructors adapt their explanations, scaffolding, and motivational support based on the learner's context: \textit{who the learner is, what they are trying to learn, and how they are engaging with the material} \cite{liu2025-learning-context,McNamara2022}. 
As large language models (LLMs) become increasingly utilized as AI tutors, a critical question emerges: \textit{Does providing learning context (LC) induce genuine pedagogical adaptation in AI instructional planning, or does it merely introduce superficial variation in responses?}

\begin{figure}[t]
    \centering
    \includegraphics[width=0.8\textwidth]{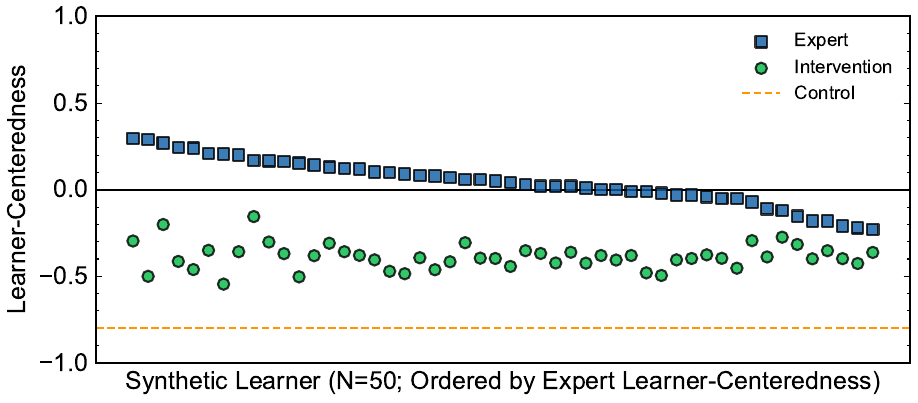}
    \caption{Learning context reduces the personalization gap. Shown are learner-centeredness scores for instructional policies produced by GPT 5.2 in control (learning context-blind) and intervention (learning context-aware) conditions, and by expert instructors across 50 synthetic learners sharing the same learning objective. A learner-centeredness greater than 0 indicates a policy that is more learner-centered while a learner-centeredness less than 0 indicates a policy that is more content-centered. Incorporating learning context consistently increases the learner-centeredness of the LLM's instructional policy relative to an objective-only baseline, but remains substantially below expert instructional planning across all learners.}
    \label{fig:adaptivity}
\end{figure}

In this paper, we evaluate personalization at the level of an LLM-tutor's \emph{instructional design policy}: the distribution of instructional strategies the model intends to prioritize in one-on-one instruction with a learner. We call this evaluation the \ppp (\pppshort): a policy-level diagnostic that uses psychometrically grounded synthetic LCs as controlled probes and expert judgments as reference signals to quantify LC-induced policy shifts and remaining personalization gaps. Using \pppshort, we measure how this policy changes when the model only has access to the learning objective versus the full LC, and quantify \emph{learner-centeredness} (i.e., the extent to which a planned policy emphasizes learner-centered strategies relative to content-centered strategies). In our case study on GPT 5.2, we see a clear trend: across 50 psychometrically grounded synthetic learners sharing the same objective, LC reliably shifts GPT~5.2 toward more learner-centered plans (Fig.~\ref{fig:adaptivity}). However, even with LC, our results indicate that the model is substantially less learner-adaptive than expert instructors, highlighting a tension: context changes what the model plans, but not enough to produce expert-like personalization.

To explore the causes of this tension, we conduct an analysis of \emph{pedagogical relevance-influence alignment}: we test which learner characteristics the model appropriately uses, neglects, or treats as spuriously influential relative to expert judgments of pedagogical relevance. This analysis reveals systematic misalignment, including neglected learner features and apparent ``hallucinated'' relevance driven by irrelevant characteristics. 

Together, our analyses of learner-personalization (Sec.~\ref{sec:intervention}) and the misalignment between pedagogical relevance and influence (Sec.~\ref{sec:misalignment}) support an operational takeaway: {\em current LLM tutors exhibit measurable adaptation to the LC, but their instructional planning remains non-expert and insufficiently robust to distracting context}. Rather than assuming that LC automatically yields effective adaptation, \pppshort supports targeted evaluation of when context helps, when it fails, and which learner characteristics should be prioritized to improve the personalization of instructional plans. Our contributions are summarized as follows:
\begin{enumerate}
    \item We introduce \pppshort, a framework for measuring personalization in LLM tutoring via changes in an instructional design policy and its learner-centeredness.
    \item Within \pppshort, we provide a method for synthetically generating psychometrically grounded LCs 
    and using them as diagnostic probes for evaluating LLM-based tutoring systems.
    \item We present a case study of GPT~5.2 that: (i) illustrates how LC improves alignment with expert planning, and (ii) diagnoses the relevance--influence misalignment that persists despite the LC.
\end{enumerate}

\noindent
By turning ``personalization'' into a measurable phenomenon and exposing where models fail to effectively utilize LC, our approach enables targeted improvements to LC-aware LLM tutoring systems before they reach real learners.

\section{Related Work}

\subsection{Personalized Learning and AI Tutoring}

A central finding in the tutoring literature is that instruction that adapts to a learner's evolving state can yield substantial learning gains relative to less adaptive instruction. Classic work framed one-to-one tutoring as an aspirational upper bound for individualized instruction \cite{bloom1984-twosigma}, while later syntheses suggest typical effects are smaller but consistently positive across both human tutoring and well-designed intelligent tutoring systems (ITS) \cite{vanlehn2011-relative,ma2014-its,kulik2016-its}. Recent policy-focused meta-analyses further suggest these impacts can attenuate when tutoring is scaled and evaluated with standardized outcomes, underscoring the importance of measuring what forms of adaptivity are actually realized in practice \cite{kraft2024-tutoring}.

With the arrival of LLMs, LLM-based tutoring systems have emerged that use natural language generation to provide conversational support \cite{nye2023-dialog-tutoring,wang2024-llm-edu-survey}. However, much of this work implicitly assumes that personalization will emerge from prompting or access to learner information, without directly measuring whether (or how) LC changes the tutor's instructional \emph{planning} decisions.

\subsection{Learner Modeling}

Learner modeling has been a core component of ITS for decades, enabling systems to represent learner knowledge, misconceptions, and other factors to drive adaptation \cite{corbett1995-kt,brusilovsky2007-user-models,conati2013-student-modeling}. Digital Promise's Learner Variability Navigator synthesizes a broad set of learner modeling literature into a framework for understanding learner characteristics across cognitive, social-emotional, and background dimensions \cite{lvn}. The Motivated Strategies for Learning Questionnaire (MSLQ) provides validated psychometric instruments for operationalizing motivational and self-regulated learning constructs \cite{pintrich1991-mslq}. Unlike traditional learner models used for real-time adaptation or mastery tracking, we use learner models here as controlled experimental instruments for probing how learner information shapes LLM instructional planning.

\subsection{Context in LLM-Based Systems}

A growing body of work studies how different forms of context shape LLM behavior, including prompt engineering \cite{white2023-prompt-patterns}, retrieval-augmented generation \cite{lewis2020-rag}, and tool-augmented reasoning/acting \cite{yao2023-react,schick2023-toolformer}. These approaches primarily emphasize task and information context (what to do and what to know), rather than learner-centered educational context (who the learner is and how instruction should adapt). The \emph{Learning Context} framework articulates LC as a structured, multi-layer construct encompassing goals, learner characteristics, instructional constraints, and environment conditions, and calls for systematic methods to evaluate how LLM systems perceive and act on such context \cite{liu2025-learning-context}. Our work directly addresses this call by isolating LC variables and quantitatively measuring their influence on instructional decision-making at the level of the tutor's strategy-selection policy.

\section{The \ppp: A Framework for Measuring Personalization in LLM-based Tutoring}
\label{sec:ppp}

In this section, we define the components of the \ppp (\pppshort): (i) how we generate controlled learning contexts as probes, (ii) the instructional design decision task and policy estimation procedure, (iii) expert annotations used as reference signals, and (iv) metrics for policy alignment and learner-centeredness.

\subsection{Learning Context Generation}

In order to synthetically generate realistic LCs, we develop a method for generating learner profiles that are psychometrically consistent with the MSLQ. We base our method on the MSLQ because it is a widely used, validated instrument for operationalizing motivation (e.g., self-efficacy, task value, test anxiety) and learning strategies (e.g., elaboration, metacognitive self-regulation, help seeking). It provides empirical means, variances, and inter-construct correlations that enable us to generate psychometrically plausible learner profiles rather than ad hoc trait lists. Additionally, the individual item prompts are well-suited for rephrasing into natural-language descriptions that can be used as LCs for an LLM.

Rather than treating learner characteristics as a flat list of attributes, we treat them as a structured, correlated latent state. This allows us to generate learner profiles that are realistic enough to be pedagogically meaningful, while remaining fully controllable for experimental analysis.


\vspace{0.5em}
\noindent
\textbf{Latent construct score generation.}
Pintrich et al. \cite{pintrich1991-mslq} report empirical estimates the mean response for each MSLQ construct, the standard deviations for the responses, and inter-construct response correlations based on a large learner sample. We model the learner profiles at the construct level as samples from a multivariate normal distribution with these empirical statistics:
\begin{equation}
    \mathbf{z} \sim \normaldist{\boldsymbol{\mu}}{\boldsymbol{\sigma} \mathbf{R} \boldsymbol{\sigma}^\top} \,,
\end{equation}
where $\boldsymbol{\mu}$ is the vector of construct means, $\boldsymbol{\sigma}$ is the vector of construct standard deviations, and $\mathbf{R}$ is the correlation matrix of the constructs. This model ensures that generated learner profiles preserve empirically observed relationships among learning constructs.

\vspace{0.5em}
\noindent
\textbf{Item response generation.}
Each MSLQ construct is measured by multiple questionnaire items with known factor loadings. We model the item responses as a linear function of the latent constructs, plus noise:
\begin{equation}
    \mathbf{x} = \boldsymbol{\rm \Lambda} \mathbf{z} + \boldsymbol{\eta} \,,
\end{equation}
where $\boldsymbol{\rm \Lambda}$ is the matrix of factor loadings with entries
\begin{equation*}
    {\rm \Lambda}_{ij} = \begin{cases}
        \lambda_i & \text{if item } i \text{ measures construct } j \\
        0 & \text{otherwise} \,,
    \end{cases}
\end{equation*}
and $\boldsymbol{\eta} \sim \normaldist{\mathbf{0}}{\boldsymbol{\rm \Sigma}}$ is item-specific noise with covariance matrix
\begin{equation*}
    {\rm \Sigma}_{ij} = 
    \begin{cases}
        \sigma_{c(j)}^2 (1 - \lambda_j^2) & \text{if } i = j \\
    0 & \text{otherwise} \,,
    \end{cases}
\end{equation*}
where $\sigma_{c(j)}$ is the standard deviation of the construct that item $j$ measures.

\vspace{0.5em}
\noindent
\textbf{Qualitative context synthesis.}
To make learner profiles usable as LC for an LLM, we convert the numeric item responses into qualitative descriptions. For each construct, we select the item with the highest factor loading and prompt an LLM to rephrase the corresponding item statement into a natural-language observation that reflects the generated score. Importantly, this step is constrained to re-expression rather than elaboration: the LLM is instructed not to introduce new information, but only to translate quantified characteristics into qualitative forms.

To complete the LC, we append a learning objective to the qualitative learner characteristics profile. Learning objectives are sourced from high-quality peer-reviewed subject matter expert (SME) authored OpenStax textbooks, specifically, the OpenStax Calculus Volume 1 book \cite{strang2016-openstax-calculus-1}. These learning objectives are rephrased as observations to maintain consistent learner-centered framing across both the learner characteristics and the learning objective. This construction allows us to systematically manipulate LC while preserving psychological realism, enabling controlled analysis of how learner information influences instructional strategy selection. Table~\ref{tab:learning-context} illustrates an example of a complete synthetic learning context.

\begin{table}[t]
    \centering
    \caption{Example of a synthetic learning context synthesized from a MSLQ profile.
    }
    {\scriptsize
    \begin{tabular}{p{2cm}p{9cm}}
        \toprule
        \textbf{Learner \newline Profile} & The learner tends to have difficulty persisting with coursework when the materials feel dull or uninteresting, and may not consistently keep working until finishing under low-interest conditions. The learner strongly believes that with sufficient effort they will be able to understand the course material. The learner only sometimes tries to decide whether a theory, interpretation, or conclusion presented in class or in readings is supported by good evidence. The learner moderately likes the subject matter of this course. The learner frequently experiences an uneasy, upset feeling when taking an exam. \\
        \midrule
        \textbf{Learning \newline Objective} & The learner wants to practice using the Fundamental Theorem of Calculus to evaluate definite integrals. \\
        \bottomrule
    \end{tabular}
    }
    \label{tab:learning-context}
\end{table}

\subsection{Instructional Design Task}
\label{sec:instructional-design-task}

To evaluate how LC affects LLMs' instructional design decisions, we formalize instructional design as a controlled decision task. We model the LLM as an instructional designer following a policy $\pi$ that maps a learner's LC to a set of instructional strategies. The components of this formalism are: (a) the instructional state space $\mathcal{S}$, which is the set of all possible LC, (b) the instructional action space $\mathcal{A}$, which is a discrete subset\footnotemark[1]{} of instructional strategies sourced from the Digital Promise Learner Variability Navigator \cite{lvn}, and (c) the instructional policy $\pi$, which is a distribution over the action space, conditioned by the state space. This formalism enables controlled policy estimation and comparison across conditions (e.g., LC-blind vs. LC-aware) and relative to a reference expert policy.

\footnotetext[1]{The instructional strategies are filtered to ensure they are implementable by an LLM in a text-based medium. The set of strategies is given in Table~\ref{tab:strategies}.}

\vspace{0.5em}
\noindent
\textbf{Policy estimation.}
We provide the LLM with a state $s \in \mathcal{S}$ and the action space $\mathcal{A}$, and we prompt it to select the top $k$ most critical actions to prioritize in 1:1 instruction with the learner described by $s$. To capture the model's underlying distribution, we repeat the task $N$ times to estimate\footnotemark[2]{} the policy $\widehat{\pi}$:
\begin{equation}
    \widehat{\pi}(a \mid s) = \frac{\mathrm{count}(a \mid s)}{\sum_{a' \in \mathcal{A}} \mathrm{count}(a' \mid s)} = \frac{\mathrm{count}(a \mid s)}{N \cdot k} \,,
\end{equation}
where $\mathrm{count}(a \mid s)$ is the number of times action $a$ is selected in state $s$ across all trials.

\footnotetext[2]{We note that the policy estimate $\widehat\pi(a|s)$ quantifies the empirical probability of \emph{planning to take} action $a$ given state $s$, and is strictly distinct from the policy $\pi(a|s)$, which quantifies the probability of \emph{executing} action $a$ given state $s$. In this work, we assume instructional execution is a causal consequence of instructional planning, and use $\widehat{\pi}$ as a proxy for $\pi$.}

\begin{table}[t]
    \caption{Instructional strategy action space (names only). We rely on LVN's standardized operational definitions \cite{lvn}.}
    \label{tab:strategies}
    \centering
    {
    \scriptsize
    \setlength{\tabcolsep}{6pt}
    \setlength{\extrarowheight}{0.8em}
    \begin{tabular}{p{0.30\columnwidth}p{0.30\columnwidth}p{0.30\columnwidth}}
    \toprule
    Accessible \newline Vocabulary and Syntax & Check-ins & Chunking \\
    Clear Structure \newline and Presentation & Competency-based Learning and Assessment & Cultural and Life \newline Experiences Connections \\
    Debriefing & Error Analysis & Explaining \newline Their Thinking \\
    Explicit Instruction & Foster Growth Mindset & Goal Setting \newline and Monitoring \\
    Guided Practice & Immediate Feedback & Inquiry-based Learning \\
    Metaphors and Analogies & Mnemonic Device & Pairing Non-examples \newline with Examples \\
    Problem-based Learning & Real-world Math & Strengths-based \newline Approach \\
    Worked Examples &  &  \\
    \bottomrule
    \end{tabular}
    }
\end{table}

\subsection{Expert Data Collection}
\label{sec:expert}

We solicited the judgments of two expert calculus instructors to (i) score the learner- vs content-centered orientation of each instructional strategy, (ii) provide a reference mapping between learner characteristics and pedagogical design choices, and (iii) rate the pedagogical relevance of each learner characteristic. The experts completed three structured annotation tasks:

\vspace{0.5em}
\noindent
\textbf{Task 1: Strategy orientation.} For each instructional strategy, experts labeled its typical emphasis as \emph{learner-centered}, \emph{content-centered}, or \emph{balanced}, and provided a 1--2 sentence justification. Experts demonstrated a reasonably strong degree of alignment in their ratings (Krippendorff's $\alpha=0.78$). These labels were coded to integer values $+1$, $-1$, and $0$, respectively, and aggregated by mean to obtain a learner-centeredness score function $f: \mathcal{A} \to [-1, 1]$. This score function is used to summarize the learner-centeredness of a given design policy (see Sec.~\ref{sec:metrics}).

\vspace{0.5em}
\noindent
\textbf{Task 2: Instructional design task.} For each learner characteristic, the experts selected the top $5$ instructional strategies they would prioritize when designing targeted calculus instruction for a student exhibiting that characteristic, with brief justifications. Across characteristics, experts' exhibited an average agreement of 2.13 out of 5 strategies. We use these selections to form individual policies for each SME, and then take the normalized sum of them to get the aggregate expert design policy used in Sec.~\ref{sec:intervention}.

\vspace{0.5em}
\noindent
\textbf{Task 3: Characteristic relevance.} Experts rated the expected impact of each learner characteristic on instructional design on a 1--5 scale (1: low impact; 5: high impact), again with brief justifications. Experts demonstrated a moderate degree of alignment in their ratings (Krippendorff's $\alpha=0.58$). These ratings were aggregated by mean to obtain the expert-defined relevance scores used in Sec.~\ref{sec:misalignment}.

\subsection{Metrics}
\label{sec:metrics}


\textbf{Policy deviation.}
We quantify differences between instructional policies using the Total Variation Distance (TVD):
\begin{align}
    \mathrm{Deviation}(\pi, \pi'; s) 
    &= \TVD\bigl(\pi(\cdot\mid s), \pi'(\cdot\mid s)\bigr) \nonumber\\
    &= \frac{1}{2} \sum_{a \in \mathcal{A}} \left|\pi(a \mid s) - \pi'(a \mid s)\right| \,,
\end{align}
where $\mathcal{A}$ denotes the instructional action space. The TVD ranges from 0 (identical policies) to 1 (maximally distinct policies). We use TVD due to its simplicity, metric properties, and stability for low-probability actions.

\vspace{0.5em}
\noindent
\textbf{Learner-centeredness.}
As a complementary measure, we quantify the degree to which an instructional policy emphasizes learner characteristics versus learning content. Given a policy $\pi$ and state $s$, we define the policy-level learner-centeredness function $F(\pi \mid s)$ as the expected learner-centeredness across the actions $a \sim \pi(\cdot\mid s)$:
\begin{equation}
    F(\pi \mid s) = \mathbb{E}_{a \sim \pi(\cdot|s)}\bigl[f(a)\bigr] \,,
\end{equation}
where $f: \mathcal{A} \to [-1, 1]$ is the action-level learner-centeredness function defined in Section \ref{sec:expert}. Positive values of $F(\pi \mid s)$ indicate a policy that is more learner-centered, negative values indicate a policy that is more content-centered, and values close to zero indicate a policy that is balanced.

\section{Case Study 1: Measuring the Effect of Learning Context on LLM-based Instructional Design}
\label{sec:intervention}

We now apply the \pppshort framework to quantify how LC changes GPT-5.2's instructional design policy and how these changes affect alignment with the SME reference policy. We test whether providing LC leads to systematic changes in the instructional design decisions made by the LLM for a given learning objective. We also investigate indications of personalized instruction, that is, whether this shift places greater relative emphasis on learner-centered strategies compared to learning content-focused strategies.

\vspace{0.5em}
\noindent
\textbf{Setup.}
We generate estimates of the LLM-based control (LC-blind) and intervention (LC-aware) instructional design policies using the instructional design task described in Section \ref{sec:instructional-design-task} with $N=100$ trials and $k=5$ top strategies selected per trial. We also retrieve an aggregated estimate of the expert instructional design policy from SMEs.

\begin{figure}[t]
    \centering
    \includegraphics[width=0.8\textwidth]{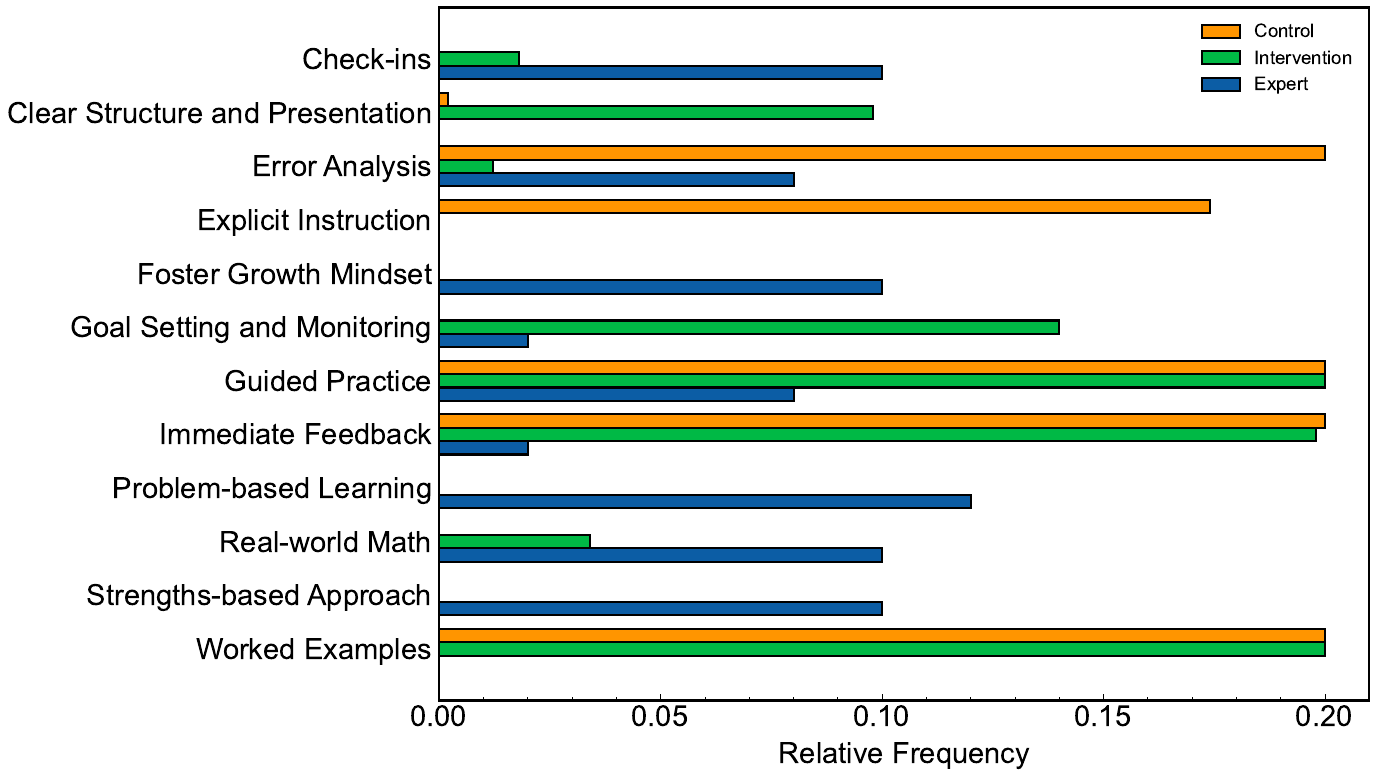}
    \caption{Influence of learning context on instructional policy estimates.
    Relative frequencies of pedagogical strategies selected by GPT 5.2 under a learning context-blind control condition, a learning context-aware intervention condition, and by expert instructors, for a learner profile characterized by moderate subject interest, high growth mindset, test anxiety, and difficulty persisting under low-interest conditions. Although inclusion of learning context shifts the LLM's instructional policy toward greater learner awareness (e.g., goal setting and monitoring), both LLM conditions remain concentrated on procedural strategies relative to the expert policy, which emphasizes a range of motivational and contextual practices. Pedagogical strategies are derived from the Digital Promise Learner Variability Navigator \cite{lvn}.}
    \label{fig:policies}
\end{figure}

\vspace{0.5em}
\noindent
\textbf{Results.}
\textit{LC shifts strategy selection toward expert planning.} Figure~\ref{fig:policies} illustrates an example of how the LLM's design policy changes when LC is available alongside the expert reference policy. In the control condition, the LLM heavily concentrates policy mass on content-focused strategies (e.g., Worked Examples, Guided Practice). When LC is provided, the policy's mass is reallocated toward strategies that consider learner characteristics and engagement (e.g., Goal Setting and Monitoring, Clear Structure and Presentation), resulting in a distribution that is qualitatively closer to the expert profile on several learner-centered dimensions. However, some expert-preferred, learner-responsive strategies remain comparatively under-selected by the LLM even with LC suggesting that the model's adaptation is limited rather than fully expert-like.

\begin{figure}[t]
    \centering
    \includegraphics[width=0.6\textwidth]{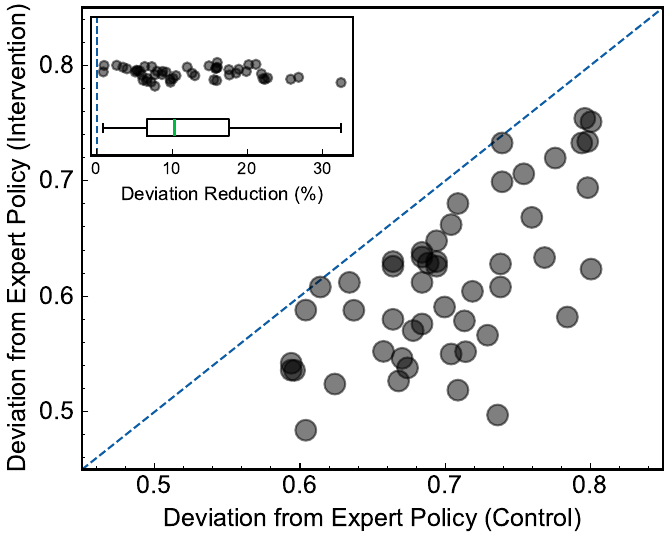}
    \caption{Comparison of LLM-expert policy deviation (measured by Total Variation Distance) under control (learning context-blind) and intervention (learning context-aware) conditions across learning contexts. A deviation of 0 indicates perfect alignment with expert instructional judgment. \textbf{(Main figure)} Points below the diagonal indicate reduced deviation from expert instructional judgment when learning context is provided. \textbf{(Inset)} The distribution of deviation reductions shows that every learning context exhibited decreased LLM-expert divergence under the context-aware condition.}
    \label{fig:deviation}
\end{figure}

\vspace{0.5em}
\noindent
\textit{LC systematically reduces the LLM's deviation from expert policy.} Across synthetic learners, LC decreases the total variation distance between the LLM policy and the expert policy: every learner falls below the identity line ($y=x$) when plotting intervention deviation against control deviation (Fig.~\ref{fig:deviation}, main). Moreover, deviation reductions are observed in every case, with more than half of LCs inducing at least a 10\% reduction in deviation from the expert policy (Fig.~\ref{fig:deviation}, inset).

\vspace{0.5em}
\noindent
\textit{LC increases learner-centered planning, but not to the level of experts.} We observe that LC shifts the LLM upward along the learner-centeredness axis toward greater learner-centered planning (Fig.~\ref{fig:adaptivity}). However, the learner-centeredness of the LLM design policy remains substantially below that of the expert policy, indicating that LC does not reliably yield expert-like personalization.

\section{Case Study 2: Diagnosing the Instructional Misalignment of LLMs}
\label{sec:misalignment}

Building on the case study in Sec.~\ref{sec:intervention}, which shows LC tends to move policies toward experts, in this section we explain where the remaining gap comes from by testing whether the learner characteristics that drive policy changes are the same ones experts deem pedagogically relevant.

\vspace{0.5em}
\noindent
\textbf{Setup.}
To estimate the influence of individual learner characteristics on instructional planning, we perform a leave-one-out analysis over a fixed learning context $s_*$ composed of characteristics $\{c_1,\dots,c_{23}\}$, which contain 15 MSLQ-sourced characteristics and 8 null characteristics (e.g., ``The learner has two siblings''). This mixture of relevant and irrelevant characteristics is designed to stress-test the influence of each characteristic on instructional planning.

For each characteristic $c_i$, we form a perturbed context $s_{-i}=s_*\setminus\{c_i\}$ by omitting that characteristic and measure its influence as the change in instructional policy, quantified by:
\begin{equation}
    \mathrm{Influence}(c_i ; \pi, s_*) = \TVD\bigl(\pi(\cdot\mid s_*),\pi(\cdot\mid s_{-i})\bigr) \,.
\end{equation}
We then compare observed influences to the expert-rated pedagogical relevance of each characteristic (see Sec.~\ref{sec:expert}) to identify (i) aligned features (high relevance, high influence), (ii) neglected features (high relevance, low influence), (iii) hallucinated relevance (low relevance, high influence), and (iv) irrelevant features (low relevance, low influence).

\vspace{0.5em}
\noindent
\textbf{Results.}
The expert-rated relevance scores confirm that null characteristics are rated as low relevance, while MSLQ-derived characteristics are rated as moderately to highly relevant. Most MSLQ characteristics fall into the aligned region, and most null characteristics correctly have relatively little influence on the LLM's design policy. However, we observe a subset of null characteristics with non-trivial influence, signifying hallucinated relevance. We also observe some highly rated characteristics with low influence, indicating neglected features and opportunities for improvement. Overall, we find no evidence of a linear relationship between pedagogical relevance and influence on LLM design policy (Spearman's $\rho\approx0.11$, $p\approx0.61$), suggesting that the model's learned parameters do not encode a reliable prioritization of learner characteristics based on their pedagogical relevance.

\begin{figure}[t]
    \centering
    \includegraphics[width=0.6\textwidth]{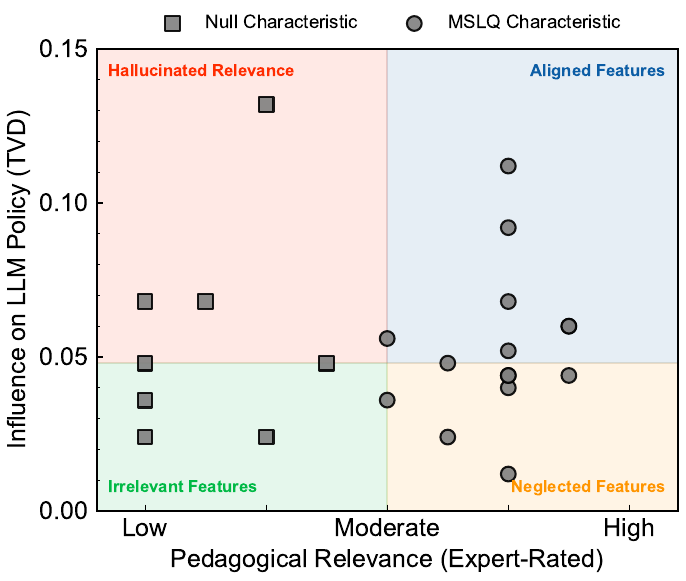}
    \caption{Relationship between pedagogical relevance of learner characteristics and their observed influence on the LLM's instructional policy (measured via total variation). Characteristics fall into four regions: aligned features (high relevance, high influence), neglected features (high relevance, low influence), hallucinated relevance (low relevance, high influence), and irrelevant features (low relevance, low influence). The presence of neglected and hallucinated characteristics reveals systematic pedagogical misalignment in how learning context is utilized during instructional planning. Quadrants are defined by median influence (horizontal) and moderate relevance (vertical).}
    \label{fig:alignment}
\end{figure}

\section{Discussion}

Our findings demonstrate that LC materially shapes instructional planning in LLM-based tutoring systems. Providing LC consistently alters the distribution of instructional strategies selected by the model and increases learner-centeredness relative to an objective-only baseline across diverse synthetic learner profiles. This indicates that LLMs are sensitive to learner information at the level of instructional planning, rather than merely producing superficial variation in responses.

However, sensitivity to LC does not guarantee pedagogically appropriate personalization. Even when LC is provided, the LLM's instructional design policies remain systematically less adaptive than those of expert instructors. This suggests personalization should be evaluated not only by whether context changes model behavior, but by whether those changes align with pedagogical intent; simply appending learner profiles to prompts can shift planning without yielding expert-like adaptation.

Our diagnostic analyses further show that learner characteristics do not exert uniform influence on planning. Some pedagogically relevant characteristics (e.g., Control Beliefs, Self-Regulation) strongly affect instructional decisions, while others experts deem important (e.g., Rehearsal, Critical Thinking) have little observable impact. Conversely, low-relevance characteristics (e.g., Preference for Physical Books, Interest in Crossword Puzzles) can still substantially influence planning, indicating miscalibrated prioritization relative to expert principles.

Taken together, effective personalization requires more than access to learner data: it requires mechanisms for determining which aspects of LC should matter for instructional planning, and how strongly. Without such mechanisms, LLM tutors may exhibit LC sensitivity that is incomplete or misaligned with pedagogical goals.

\vspace{0.5em}
\noindent
\textbf{Limitations.}
While the \pppshort framework provides a principled lens for evaluating the impact of LC on LLM-based instructional planning, several limitations must be acknowledged. First, we note that the leave-one-out analysis examines marginal impact. In reality, learner characteristics often interact in complex, non-linear ways that are not fully captured by this analysis. Second, the results are specific to one domain (calculus). It remains to be seen whether the same patterns hold for other domains, especially more subjective or qualitative domains like history or literature. Third, we have evaluated instructional design, i.e., the model's intent. However, we identify that actions are not necessarily causal consequences of intents, so in practice the LLM may have an \textit{execution policy} and a \textit{design policy} that are distinct. These limitations can be addressed in future work, including broader domains, as we discuss in the next section.

\vspace{0.5em}
\noindent
\textbf{Future Work.}
Several avenues for future research emerge from this work. Extending the analyses to additional instructional domains would help clarify the generality of observed context effects and diagnostic patterns. More expressive experimental designs, such as factorial manipulations of learner characteristics, could illuminate interaction effects that are not captured by marginal analyses.

From a system-design perspective, our results motivate the development of pedagogical prioritization frameworks that explicitly encode which learner characteristics should influence instructional design and under what conditions. Such frameworks could inform prompt design, model fine-tuning, or hybrid systems that combine LLMs with structured learner models. Finally, closed-loop evaluations that compare LC-aware and LC-blind tutoring systems on learning outcomes would provide an important complement to the design-level analyses presented here.

\section{Conclusion}

In this paper, we introduced the \ppp (\pppshort), a policy-level diagnostic that uses psychometrically grounded synthetic LCs to measure how learning context shifts an LLM's instructional strategy distribution relative to expert planning. With \pppshort, we showed that LC induces emphasis on learner-centered strategies, yet substantial personalization gaps and relevance--influence misalignment persist, including neglected pedagogically relevant traits and spurious sensitivity to irrelevant characteristics. By turning personalization into a measurable object and revealing where LC use fails, our approach enables targeted improvements and more responsible evaluation of LC-aware LLM tutors.





\begin{credits}
\subsubsection*{\ackname}
This work was supported by ONR grant N00014-23-1-2714, ONR MURI N00014-20-1-2787, DOE grant DE-SC0020345, and DOI grant 140D0423C0076.

\subsubsection*{\discintname}
The authors have no competing interests to declare that are relevant to the content of this work.
\end{credits}

\bibliographystyle{splncs04}
\bibliography{refs}

\end{document}